\documentclass[amstex,epsf,amssymb,usenatbib]{mn2e}

\usepackage{epsfig}
\usepackage{amsmath,amssymb}
\newcommand{\Msolar}{\mbox{\,$\rm M_{\odot}$}}        

\title{Fe-bump instability: the excitation of pulsations in subdwarf B
and other low-mass stars}

\author[C.S. Jeffery \& H. Saio]
       {C.S. Jeffery\thanks{E-mail: csj@arm.ac.uk}
\& H. Saio\thanks{E-mail: saio@astr.tohoku.ac.jp} \\
Armagh Observatory, College Hill, Armagh BT61 9DG, Northern Ireland\\
Astronomical Institute, School of Science, Tohoku University, Sendai 980-8578, Japan\\
}

\date{Accepted .....
      Received ..... ;
      in original form .....}

\pagerange{\pageref{firstpage}--\pageref{lastpage}}
\pubyear{2006}

\begin{document}

\maketitle

\label{firstpage}

\begin{abstract}
\noindent We consider the excitation of radial and non-radial
oscillations in low-mass B stars by the iron-bump opacity
mechanism. The results are significant for the interpretation of
pulsations in subdwarf B stars, helium-rich subdwarfs and extreme 
helium stars, including the EC14026 and PG1716 variables. 

We demonstrate that, for radial oscillations, the driving mechanism becomes effective
by increasing the contrast between the iron-bump opacity and the
opacity from other sources. This can be achieved either by
increasing the iron abundance or by decreasing the hydrogen abundance. 
The location of the iron-bump instability
boundary is found to depend on the mean molecular weight in the envelope
and also on the radial order of the
oscillation.  A bluer instability boundary
is provided by increasing the iron abundance alone, rather than the
entire metal component, and is required to explain the
observed EC14026 variables. A bluer
instability boundary is also provided by higher radial order
oscillations. Using data for observed and
theoretical period ranges, we show that the coolest EC14026 
variables may vary in the fundamental radial mode, 
but the hottest variables {\it must} vary in modes of higher radial order.
 
In considering non-radial oscillations, we demonstrate that g-modes
of high radial order and low spherical degree ($l<4$) may be excited in some
blue horizontal branch stars with near-normal composition
($Z=0.02$). Additional iron enhancement extends the g-mode instability 
zone to higher effective temperatures and also creates a p-mode
instability zone. The latter is essentially contiguous
with the radial instability zone. With sufficient iron, the p-mode and
g-mode instability zones overlap, allowing a small region where the
EC14026 and PG1716-type variability can be excited
simultaneously. The overlap zone is principally a function of
effective temperature and only weakly a function of luminosity. 
However its location is roughly 5000{\rm K} too low compared with the
observed boundary between EC14026 and PG1716 variables. 
The discrepancy cannot be resolved by simply increasing the iron abundance.
\end{abstract}


\section{Introduction}              
\label{intro}

The discovery of a significant contribution to stellar opacity from iron-group elements 
at temperatures around 200\,000\,K \citep{OPAL92, OP95} had major consequences for studies 
of pulsation in stars. First, it solved the long-standing Cepheid mass
problem \citep{Mos92,Kan94}. 
Second, it provided a natural explanation for hitherto 
unexplained stellar variability, notably with respect to the $\beta$ Cepheids \citep{Dzi93} 
and some extreme helium stars \citep{Sai93}. Third, it initiated searches for variability
(predicted and observed) in stars hitherto thought to be stable. 
There are now known to be several classes of variable in which pulsations are driven by the 
opacity (or $\kappa-$) mechanism excited by the Fe- or Z- opacity
bump. It is one goal of this paper to explore some of the connections
between these groups and hence to understand better the nature of
Fe-bump driven pulsations. 

\begin{figure}
\begin{center}
\epsfig{file=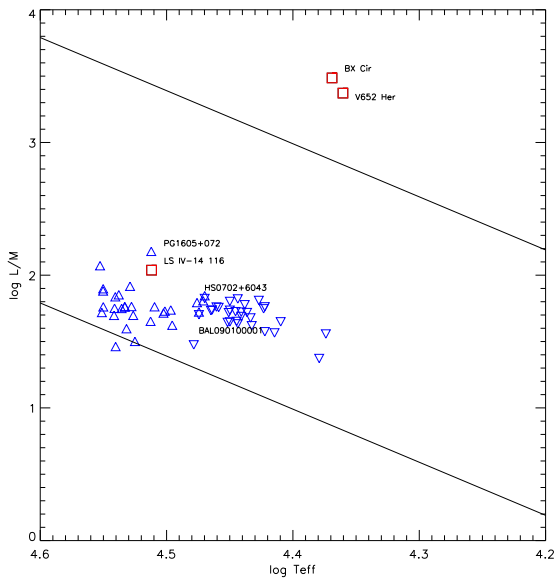,width=8cm,angle=0}
\caption[Fe-bump variables]
{The loci of low-mass variable stars believed to pulsate due to Fe-bump
  instability, including EC14026 variables (upright triangles),
PG1716 variables (inverted triangles) and
helium-rich variables (squares). A selection are labeled. HS0702+6843
and Balloon 090100001 exhibit both EC14026 and PG1716-type behaviour \citep{Sch06,Ore05}. 
Straight diagonal lines represent $\log g=6.0$ (below) and 4.0
(above). 
These features are reproduced in Figs.~\ref{Z-stable} to
\ref{Fe-modes}. }
\label{sdbv_obs}
\end{center}
\end{figure}

An early success was the explanation of radial pulsations in the
extreme helium star V652\,Her \citep{Sai93}, together with the
prediction \citep{Sai94} and subsequent discovery of radial pulsations
  in a second helium star, BX\,Cir \citep{Kil95}. In these stars, both
  with periods of $\sim 9\,300$s and with masses 0.59 and 0.47 \Msolar\ 
  respectively \citep{Jef01,Woo02} (Fig.~\ref{sdbv_obs}), the
  cause of pulsation instability is clearly the Fe- opacity bump, exaggerated
  by a reduction of the background opacity due to hydrogen at
  these temperatures. The failure to detect any variability in a third
  helium 
  star, HD144941, of similar temperature and
  luminosity but very low metallicity
  \citep{Jef96,Jef97}, confirmed the r\^ole of the Fe-opacity in driving
  pulsations in these low-mass early-type stars. 

Another notable success has been the prediction and discovery of short-period 
pulsations in subluminous B (sdB) stars \citep{Cha96, Cha97, Kil97, Bil97}. 
Briefly, multi-periodic oscillations with periods between 90 and 600 seconds are observed in 
approximately 10\% of hot subdwarf B stars; such
stars are variously known as EC14026 variables (after the class prototype EC\,14026--2647) or sdBVs. 
These pulsations are successfully explained if the stars are ``extreme horizontal branch stars'', 
that is they consist of a helium-burning core of some 0.5\Msolar\ overlaid by a very thin layer of hydrogen. 
Because of the high effective temperature and the high surface
gravity, theoretical models have shown that  radiative forces on the ions 
in the stellar envelope act differentially such that substantial chemical gradients are established 
over a diffusion time scale $\sim 10^5$y \citep{Mic89}. The consequent levitation and accumulation 
of iron in layers at around  200\,000\,K \citep{Cha95} enhances the Fe-opacity bump sufficiently that 
radial and non-radial p-mode oscillations are excited. The theory
successfully explained the observed distribution  (Fig.~\ref{sdbv_obs}) of pulsating 
and non-pulsating sdB stars in effective temperature and surface gravity \citep{Cha01}. 
Although consequential more on frequency eigenvalues than 
  stability criteria, the theory is also  sufficiently well developed that it has been possible to
 compare predicted and observed pulsation frequencies in some pulsating sdB stars \citep{Bra01, Cha05a, Cha05b}. 

The unexpected discovery of oscillations with periods of a few hours in many sdB stars lying red-ward 
of the EC14026 instability domain presented a new challenge \citep{Gre03}. While radiative levitation of iron
is still operative in these stars, sometimes known as PG1716 variables (after the prototype PG\,1716+426), 
p-modes are reported to be stable in the chemically stratified models
\citep{Cha01}. On the other hand, non-radial g-modes of high radial order 
($k\geq10$) and high spherical degree ($l\geq3$) were found to be unstable, but only in models of stars cooler 
than those in which variability had been detected \citep{Fon03}. 
While the observed periods imply a g-mode origin, modes of such high degree 
are not  generally thought  to be observable as variations in total flux due to geometric cancellation.
The challenge is therefore to shift the g-mode blue edge to higher effective temperatures and 
to excite modes of lower spherical degree. 

A further challenge has been provided by the detection of variability in the helium-rich 
sdB star LS\,IV$-14^{\circ}116$ with periods of $\sim1800 - 3000$s
\citep{Ahm05}. While the canonical sdB stars mostly have surface helium
abundances one tenth of the solar value ($\sim10\%$ by number), the
much rarer He-rich sdB stars comprise a spectroscopically distinct and
highly inhomogeneous group with surface helium abundances ranging from 
some 30\% to nearly 100\% by number, and surface gravities over a much
broader range than seen in normal sdB stars \citep{Ahm03}. 
The natural response to the discovery of
pulsations in LS\,IV$-14^{\circ}116$ was to try to explain them in
terms of Fe-bump instability in either a helium star like V652\,Her,
or in some sort of mutant sdB star.  
With $T_{\rm eff}$ similar to the EC14026 
stars, but with a lower surface gravity (higher $L/M$ ratio), the
pulsation periods were too long to be explained by Fe-bump driven
p-modes \citep{Ahm05}. One question is whether g-modes might be excited in such a star.
This  challenge is important  because a viable picture of the star that explains both the $L/M$ ratio 
and the high surface abundance of helium has yet to be established. Any model that successfully
reproduced the observed oscillations would assist this process.

In order to address these questions, we need to develop our understanding of instability 
in low-mass stars. By gaining a general insight into what affects the type of pulsation that can be driven, 
it should be possible to determine what model properties need to be modified in order to reproduce observed 
oscillations. 

The important theme is clearly Fe-bump instability. We have already investigated the r\^ole of 
envelope hydrogen abundance in the excitation of Fe-bump pulsations \citep{Jef98}. We recall 
that pulsational instability exists in all stars of sufficiently high $L/M$ ratio due to the presence of 
strange-mode instability \citep{Gau90}. Fe-bump instability sets in
for lower $L/M$ stars with $T_{\rm eff}\sim 20 - 30\,000$K when either the 
iron-abundance is raised sufficiently or the hydrogen abundance is reduced sufficiently. The second occurs 
because reducing the hydrogen opacity increases the contrast between the iron opacity and the background 
opacity due to other sources, effectively increasing the opacity gradients 
$\kappa_T = (\partial\ln\kappa/\partial\ln T)_\rho,
\kappa_{\rho}=(\partial\ln\kappa/\partial\ln\rho)_T$ 
in the driving zone. 

Therefore there should be a physical connection between, for example, the radial pulsations seen in 
the extreme helium star V652\,Her (essentially normal iron, reduced hydrogen) and the p-mode oscillations
in EC14026 variables (normal hydrogen, enhanced iron). Note that radial modes are special cases of 
p-modes with spherical degree $l=0$; such modes are generally present in EC14026 stars alongside modes of 
higher spherical degree. However, the EC14026 instability region reported by \citet{Cha01}
lies considerably blue-ward of the Fe-bump instability finger described by \citet{Jef98}. 
Understanding this paradox represents an important step towards understanding 
Fe-bump instability in general. 

Our initial goal is therefore to explore Fe-bump instability for radial modes in comparatively simple 
(i.e. homogeneous) stellar envelopes as a function of composition
(section II). With success here, it becomes possible to restrict the volume of 
model space required to explore the stability of modes of higher
degree, in particular g-modes (section III), and also to
identify what is required in more detailed models in order to reproduce the observations accurately.

\section{Radial modes}              
\label{radial}

Linear non-adiabatic models of radial pulsations have been constructed following the method 
described by \citet{Sai95}, but including the most recent OPAL95
opacities \citep{Igl96}.  In all cases the equation of state (EOS) treats
hydrogen, helium and a representative ``metal'', namely carbon. The
contribution to the EOS of varying the heavy element distribution within ``Z'' is
negligible at temperatures above 10\,000\,K since both the electron density and
the mean atomic weight are completely dominated by hydrogen and helium. 

The parameter range considered is $M/\Msolar=0.4-0.9$, $\log T_{\rm eff}/K=4.20-4.60$ and
$\log L/M=0.4-3.8$ (solar units). In terms of composition, the hydrogen mass fraction $X=0.0 - 0.9$ includes 
extremely helium-rich, normal and helium-depleted mixtures. 
The treatment of metal mass fraction $Z$ is discussed below. The choice of $L/M$ rather than $L$
as independent variable was made because pulsation properties scale roughly with $M$ (actually $M^{1/2}$), 
simplifying the generation of model grids and the inter-comparison of results for different masses.

For each combination of $M, T_{\rm eff}, L/M, X$ and $Z$, a homogeneous static stellar envelope is 
integrated assuming a standard surface boundary condition. This integration is terminated when the
total pressure exceeds $10^{17}$ dyn or the mass variable $<0.6M$, whichever occurs first.
This limit is justified for a stability analysis of radial modes (and p-modes in general)
since amplitudes decay exponentially towards the stellar center. However, the accurate 
prediction of periods for asteroseismology (for example) must use whole-star models 
so as to account for higher order effects. 
 
The first several eigenfrequencies are located and stored, including the real and imaginary components
$\omega_{\rm r}$ and $\omega_{\rm i}$,   
the period $\Pi$ and 
the number of nodes in the eigensolution. So far, it has not been necessary to analyze more than the 
first ten eigenfrequencies; the maximum number of unstable modes located so far was nine.  The
results have been collated as diagrams showing 
the stability of one or more modes and the number of unstable modes as a function of
$T_{\rm eff}$ and $L/M$. Stability is established by the sign of $\omega_i$ (negative for unstable 
modes). Results for varying $X$ and $Z$ are compared in a mosaic of similar such diagrams, on 
which are superposed the loci of selected groups of variable star. 

When identifying eigenfrequencies it is usual to identify the fundamental, first overtone, second overtone, etc,
in sequence for each model envelope, as verified by the number of nodes in the eigensolution. However, 
in the most luminous models, strange mode pulsations are excited and the one-to-one correspondence between the
ordinal number of the eigensolution identified and the number of its radial nodes frequently breaks down.
The physics of this question are discussed in more detail elsewhere \citep{Sai98}; it has no effect on the results 
presented here.

\begin{figure*}
\begin{center}
\epsfig{file=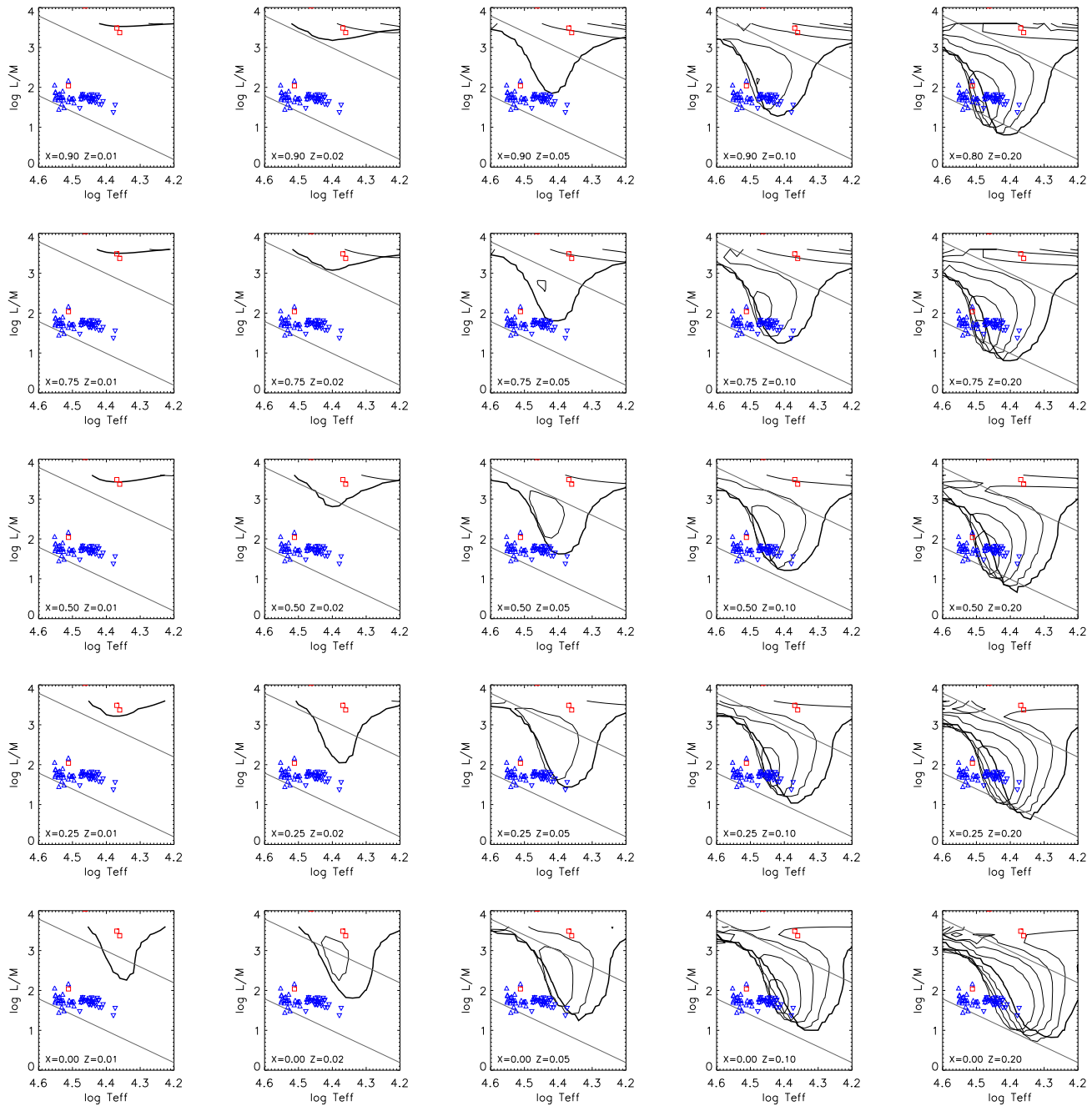,width=18cm,angle=0}
\caption[Instability Boundaries: Z]
{Instability boundaries for radial pulsation modes in stars with $M=0.5\Msolar$ and homogeneous envelopes with 
$Z = 0.01 - 0.20$.
The bold line shows the instability boundary for the longest period radial mode, normally the
fundamental $k=0$. 
Thin contours show successive boundaries for $k=1,2,3$, where these
exist. Other features as in Fig.~\ref{sdbv_obs}.
}
\label{Z-stable}
\end{center}
\end{figure*}

\begin{figure*}
\begin{center}
\epsfig{file=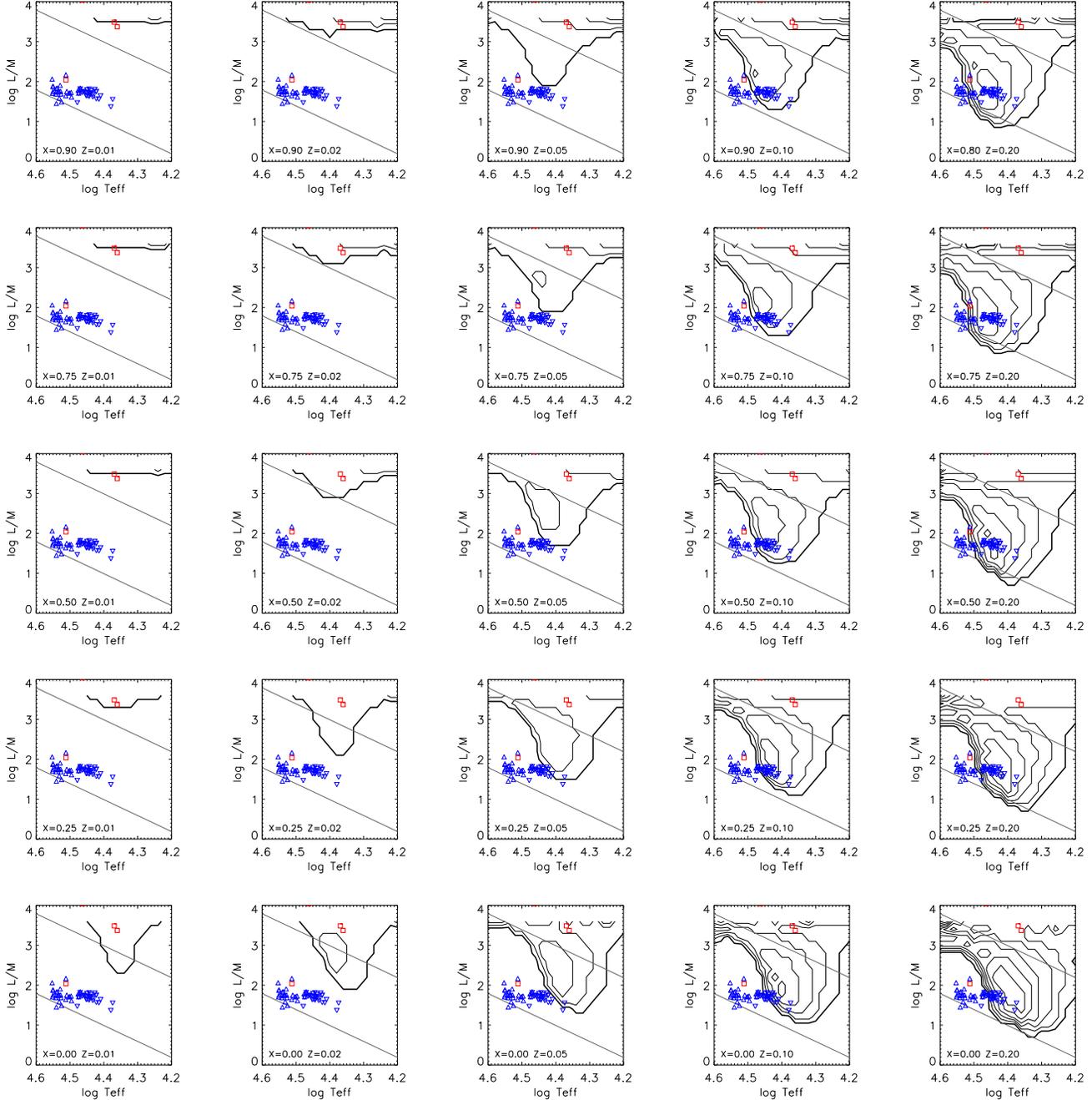,width=18cm,angle=0}
\caption[Unstable  Modes: Z]
{The number of unstable radial pulsation modes in stars with $M=0.5\Msolar$ and homogeneous envelopes with 
$X = 0.0 - 0.9$ and 
$Z = 0.01 - 0.20$.
The bold line shows the overall instability boundary; within this contour at least one radial mode is excited.
Each subsequent contour (thin lines) represents one additional
unstable mode.  Other features as in Fig.~\ref{sdbv_obs}.
}
\label{Z-modes}
\end{center}
\end{figure*}

\begin{figure*}
\begin{center}
\epsfig{file=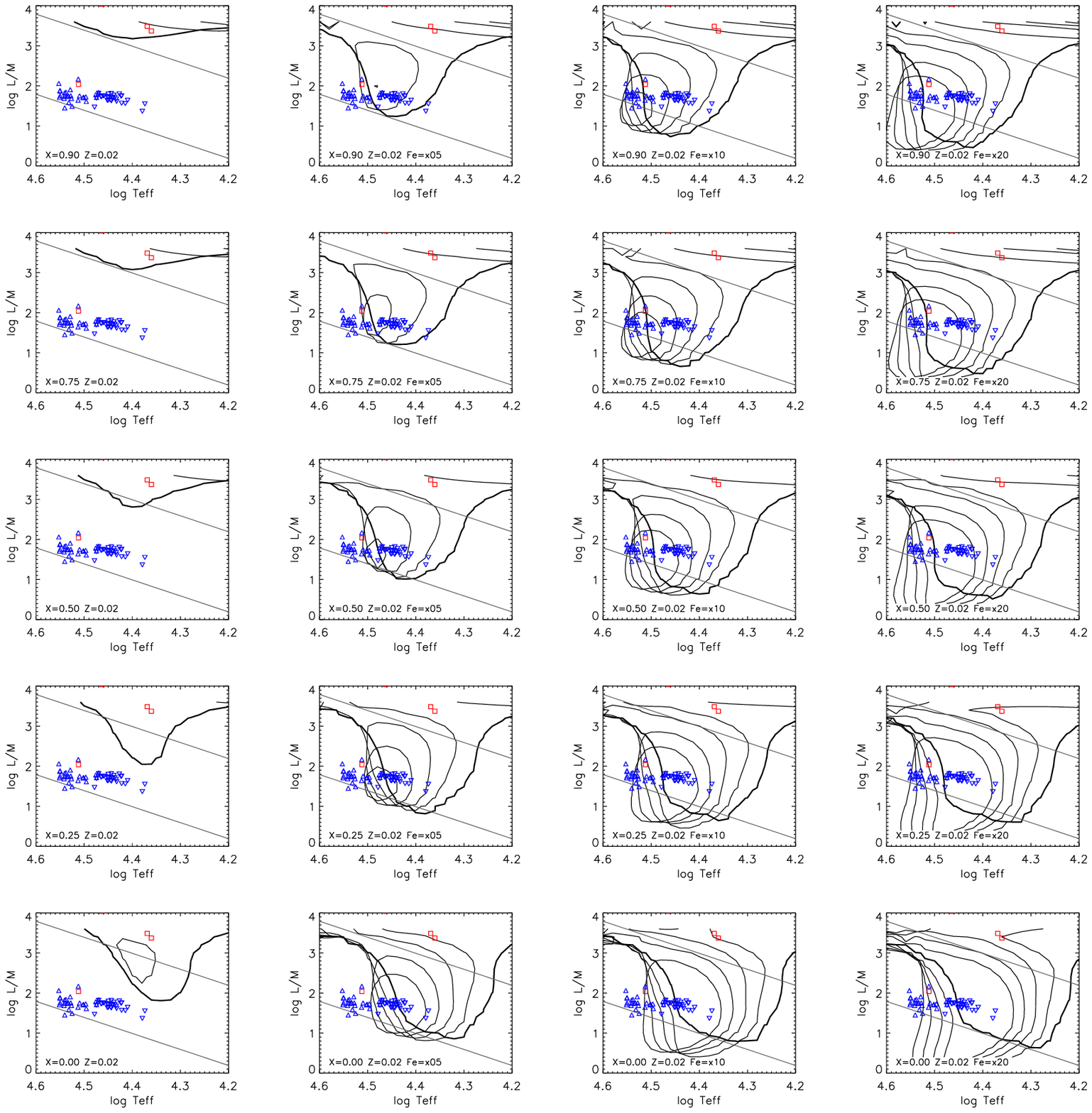,width=18cm,angle=00}
\caption[Instability Boundaries: Fe]
{As Fig.~\ref{Z-stable} but with 
$Z = 0.02$ and Fe enhanced by factors of 1, 5, 10 and 20.
}
\label{Fe-stable}
\end{center}
\end{figure*}

\begin{figure*}
\begin{center}
\epsfig{file=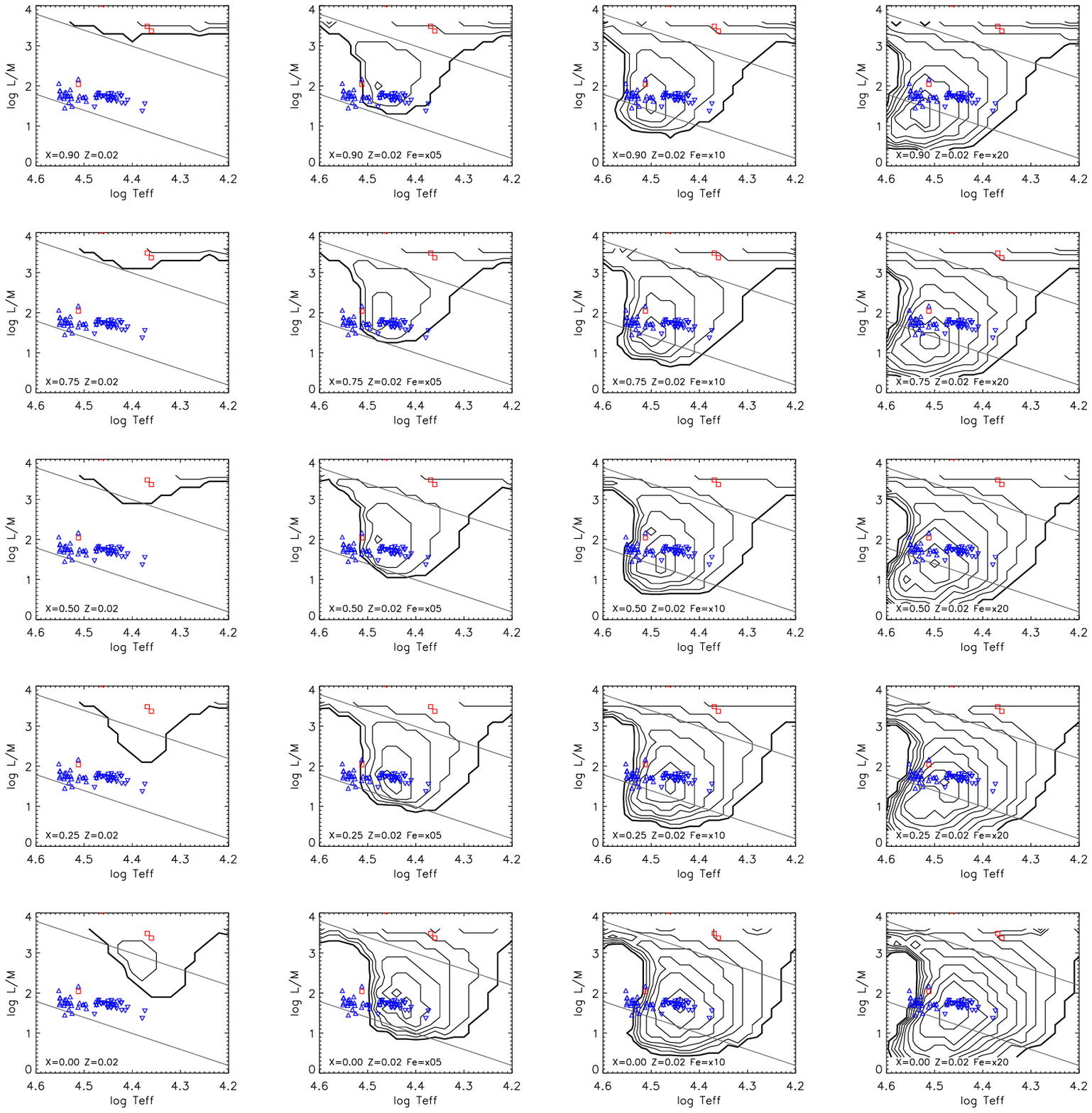,width=18cm,angle=0}
\caption[Unstable Modes: Fe]
{As Fig.~\ref{Z-modes} but with 
$Z = 0.02$ and Fe enhanced by factors of 1, 5, 10 and 20.
}
\label{Fe-modes}
\end{center}
\end{figure*}

\subsection*{Z variation} 

The first experiment was carried out on the basis that increasing the overall metallicity is known to increase
the size of the Fe-bump instability. Calculations were carried out for $Z = 0.01, 0.02,
0.05, 0.10$ and 0.20. $X$ was reduced to 0.80 when $X+Z > 1$
would otherwise have resulted. This is essentially a repeat of the
experiment described by \citet{Cha01} but over a much larger volume of
parameter space.

Figures \ref{Z-stable} and \ref{Z-modes} show how the resulting instability 
boundaries behave as functions of $X$ and $Z$ for a mass of 0.50\Msolar. The results for other masses (0.4, 
0.7 and 0.9 \Msolar) are similar. The discrete model grid is reflected in the contour shapes. 
Note that all the panels have hydrogen abundance increasing from bottom to top and metallicity 
(in whatever form) increasing from  left to right. 

For clarity, Fig. \ref{Z-stable} shows the instability boundaries for only the radial fundamental 
and the first, second and third radial overtones ($k=0,1,2,3$). 
It demonstrates a number of important general features. \\
1) The Fe-bump instability finger is not significant for normal abundances ($X=0.75, Z=0.02$). \\
2) The size of the instability finger increases with increasing $Z$ and with decreasing $X$. \\
3) For increasing $Z$, the {\it number} of excited nodes increases. \\
4) The  Fe-bump instability region, including the blue edge, becomes bluer for 
higher order modes (a similar result holds for the classical He{\sc ii} instability strip). \\
5) The blue edge for any given mode ({\it e.g.} the fundamental) is not shifted significantly 
by increasing Z, while the red edge is shifted red-ward. \\
6) The low-luminosity end of the instability finger is populated by the highest number of excited modes. \\
7) Unstable modes with $L/M\geq3.5$ are primarily strange modes. \\ 

Figure \ref{Z-modes} presents an alternative view of the 
same results but, because it includes all the excited modes up to $k=10$, 
it reflects the extent and mode-density of the instability region more effectively. 
Many of these features were already well known from previous studies \citep{Jef98,Cha01}, but it is helpful 
to reiterate them here in one unified picture.

In terms of matching the observations of 
the EC14026 stars and objects like LS\,IV$-14^{\circ}116$, this figure shows that at these high
$T_{\rm eff}$, radial modes 
(and hence p-modes in general) are not excited, even for very high $Z$
\citep[cf. Fig.3 in ][]{Cha01}.
 Despite the excitation of higher-order modes with slightly bluer blue edges, the extension to the instability
zone is insufficient.

This has a natural explanation. In attempting to shift the instability boundary, $Z$ has been 
increased in order to increase the driving by the Fe-opacity bump. At
the same time the mean molecular weight of 
the gas and the mean opacity in the layers above the driving zone has been increased. 
Therefore the Fe-bump will be located 
correspondingly closer to the stellar surface. In order to make the
blue edge bluer, it is necessary to maintain the contribution to
driving from iron without increasing the mean molecular weight and opacity in the layers above the driving layer. 


\subsection*{Fe variation}

The logical alternative is therefore to increase the iron abundance {\it without} increasing the
abundances of other heavy elements. Using the OPAL opacity table generator \citep{Igl96}, 
we constructed new sets of opacity tables in which the iron abundance alone was increased by factors 
of $f = 5, 10$ and 20, whilst the remaining elements had abundances corresponding to a $Z=0.02$ mixture.  
There is a modest effect on the overall $Z$ such that $Z=0.02+(f-1)Z_{\rm Fe}$ where $Z_{\rm Fe}=0.02188$ 
represents the mass fraction of iron within the standard heavy element mixture \citep{Gre93}.
A similar set of models was run for these mixtures. These are compared with the reference set
$Z=0.02$ in Figs.~\ref{Fe-stable} and \ref{Fe-modes}.

It is immediately apparent that increasing the iron abundance alone by a factor five
produces a predictable effect similar to increasing $Z$ from 0.02 to 0.10, but with
the major difference that the blue edge has become bluer and more vertical (in the $L/M-T_{\rm eff}$ 
diagram). As the iron abundance is further increased ({\it cf.} $X=0.75, Z=0.02, f=10$), 
a region develops which contains several excited modes and which is bluer than the Fe-bump 
instability finger seen when the entire $Z$-component  varies together. 

This is precisely the phenomenon identified by \citet{Cha97}. The overall extent of our instability 
zone is larger than identified before; the earlier calculations benefit from obeying a total iron 
conservation law by using an iron distribution obtained from a self-consistent calculation  
of radiative levitation. 

The significance here is that earlier results are reproduced in character,
if not in detail, using a simple model. It tells us that the principal characteristics of
p-mode instability in EC14026 stars may be derived from a homogeneous stellar envelope
with a factor 10 enhancement in iron abundance (assuming $Z\sim0.02$).
This provides useful pointers for a subsequent investigation of g-mode instability. 

It is interesting that reducing hydrogen concentration extends the instability finger 
towards lower $L/M$ and {\it lower} $T_{\rm eff}$. Whether this region of parameter
space is populated  by any real  stars is a matter for conjecture. If they do,  
they are likely helium- and iron-rich white dwarfs with periods of $\sim$ seconds. 
However, diffusion theory suggests that in stars of such high gravity, iron would settle 
out of the layers where it would be required to excite pulsations.  

To understand the phenomenon we recall that the instability blue edge
occurs when the thermal timescale 
\begin{equation}
\tau_{\rm th}=c_{\rm  v} T \Delta m/ L 
\label{eq_tauthdef}
\end{equation} 
becomes too small compared with the pulsation period, where $T$ is the temperature in the excitation
layer, $c_{\rm v}$ is the specific heat 
per unit mass at constant volume, and $\Delta m$ is the mass lying above the excitation layer
\citep[see][Ch. 10 for details]{Cox80}.
Let us assume an ideal gas and that the opacity in the layers above the excitation zone is expressed as 
\begin{equation}
\kappa=\kappa_0\rho^aT^{-b}
\label{eq_kappa}
\end{equation} 
with constants $\kappa_0$, $a (0\le a \le 1)$, and $b$.
Then,  we can integrate approximately, from the stellar surface to the excitation zone,  
the hydrostatic equation and the equation of diffusive flow of radiation
to obtain
\begin{equation}
\Delta m \propto (\kappa_0 L \mu^a M^a)^{-{1\over 1+a}} R^4,
\label{eq_deltam}
\end{equation} 
where $\mu$ is the mean molecular weight and $R$ is the stellar radius.
Substituting equation (\ref{eq_deltam}) into equation (\ref{eq_tauthdef}), and taking into account
$c_{\rm v} \propto 1/\mu$, we have
\begin{equation}
\tau_{\rm th} \propto \left(\kappa_0\mu^{1+2a}\right)^{-{1\over 1+a}}
(L/M)^{a\over 1+a} T_{\rm eff}^{-8}.
\label{eq_tauth}
\end{equation}
The strong dependence of $\tau_{\rm th}$ on $T_{\rm eff}$ is responsible for the blue and
the red edges of instability range.
Since the $\kappa$-mechanism is optimal when the period of pulsation is comparable to
$\tau_{\rm th}$, the instability range of a short (long) period mode is hotter (cooler)
as can be seen in Figs.~\ref{Z-stable} and \ref{Fe-stable}.
  
For a given L/M ratio, the blue-edge can be shifted blue-ward 
by increasing $\tau_{\rm th}$, {\it i.e.} by reducing $\kappa_0$ and the mean molecular weight
$\mu$. The former can be achieved by reducing $Z$  
without significantly reducing the iron abundance.
The mean molecular weight is reduced by increasing $X$. The effect is somewhat complicated 
because an increase in $X$ tends to increase $\kappa_0$.
Figs.~\ref{Z-stable} and \ref{Fe-stable} seem to indicate that the effect of reducing 
the mean molecular weight exceeds the effect of increasing opacity.

\begin{figure*}
\begin{center}
\epsfig{file=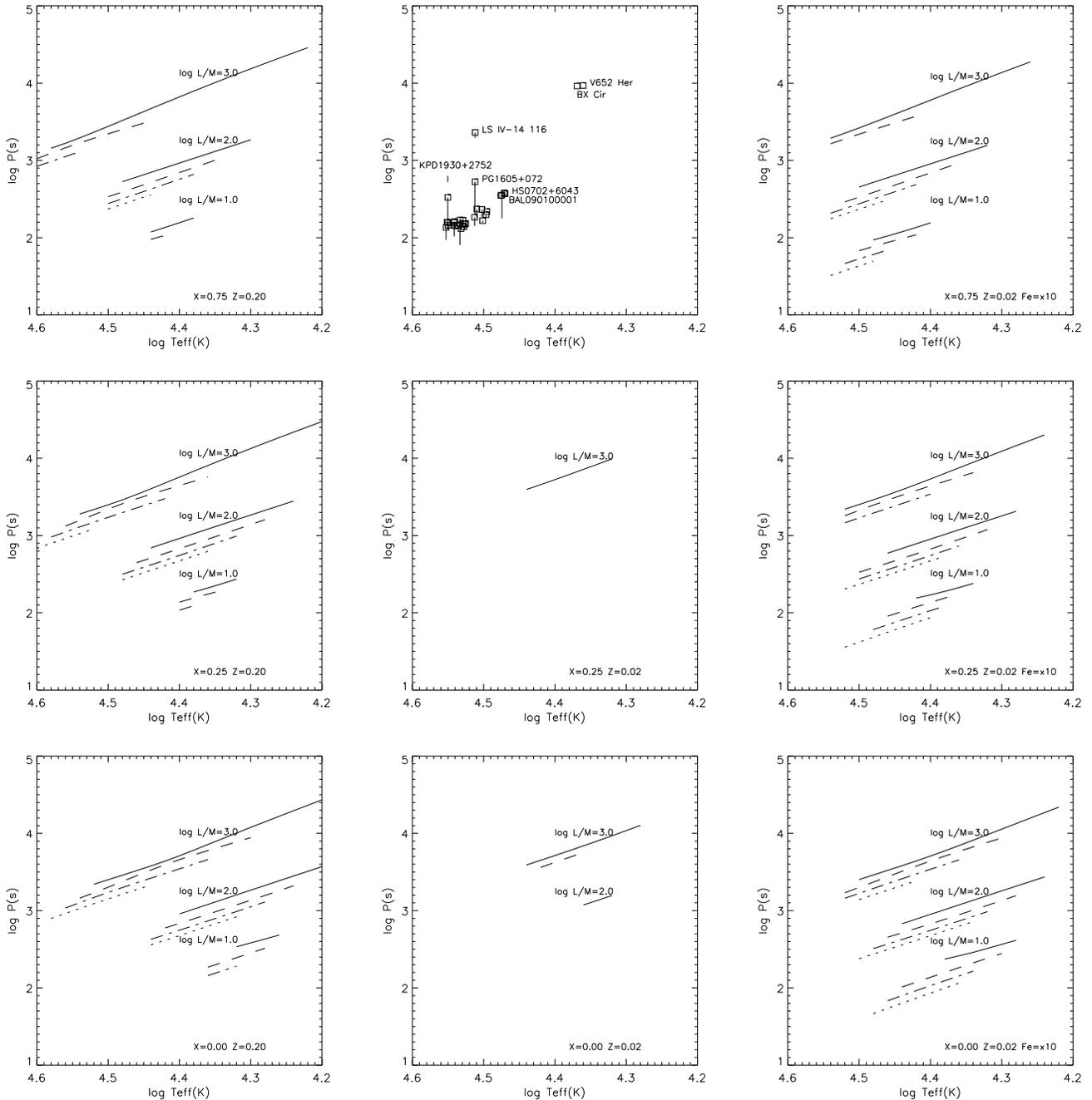,width=18cm,angle=0}
\caption[Periods for Unstable Modes]
{Periods of unstable modes due to Fe-bump instability for stars with 
$M=0.50\Msolar$ and $\log L/M = 1, 2$ and 3. Each panel shows the
  fundamental (solid), first (dashed), second (dot-dashed) and third
  (dotted) harmonics. The middle column shows models with a standard
  metal content ($Z=0.02$) and with $X$ increasing upwards (there are
  no unstable modes for $X=0.75, Z=0.02$ at the $L/M$ ratios shown). 
  The left hand column shows models with enhanced $Z=0.20$, 
  the right hand column shows models with enhanced iron only ($f=10$).
  Both columns show models with $X=0.0, 0.25, 0.75$ increasing upwards.
  The loci of selected pulsating stars are shown in the centre - top
  panel; multi-periodic stars are shown by their longest period
  (square) and their range (line). The selection includes pulsating sdB
  stars and extreme helium stars. 
}
\label{Periods}
\end{center}
\end{figure*}

\begin{figure}
\begin{center}
\epsfig{file=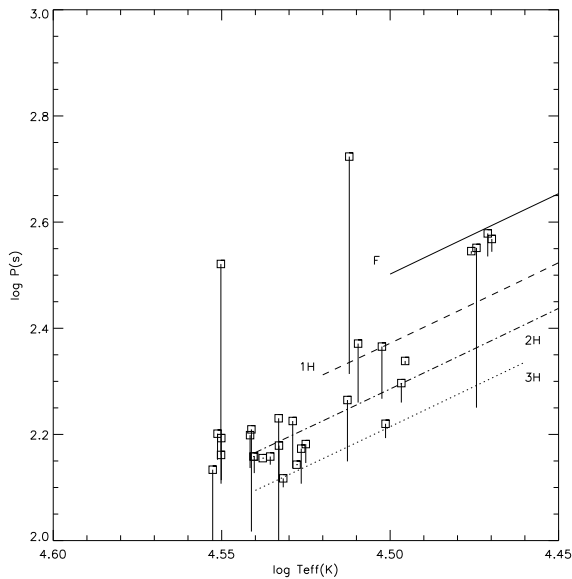,width=8cm,angle=0}
\caption[Expanded Periods]
{Fig.~\ref{Periods} redrawn with the observed periods of EC14026
  stars over-plotted on the theoretical periods of unstable radial modes 
  for $M=0.50\Msolar$, $\log L/M = 1.8$, $X=0.75, Z=0.02$ and enhanced
  iron ($f=10$). The modes are labeled by radial degree. 
}
\label{Blowup}
\end{center}
\end{figure}

\subsection*{Periods}

Predicted pulsation periods for unstable fundamental radial modes, and
for first, second and third radial harmonics are shown in
Fig.~\ref{Periods} for a range of compositions and for $\log L/M=1,
2$ and 3. The
central column represents models with a standard metallicity and
reduced hydrogen abundance, while
the left and right columns represent a uniform metal enhancement and a
selective iron enhancement respectively. Note again that the lower
$L/M$ pulsators for a selective iron enhancement are bluer than for a
uniform enhancement, and also that higher-order modes are
preferentially excited in bluer pulsators. 

For comparison, the top centre panel shows the pulsation periods and
period ranges 
for several EC14026 and extreme helium-star pulsators. A consequence of the models is that 
bluer EC14026 stars should show higher harmonics and, conversely,
there should be fewer hot sdBVS pulsating in the radial fundamental
mode. This is clearly reflected by the mean slope of the $\log P(\log
T_{\rm eff})$ relation for EC14026 stars (excluding PG\,1605+072 and 
KPD\,1930+2752) being steeper than that for a single
mode. Fig.~\ref{Blowup} compares the observations for EC14026
variables with pulsation periods for unstable radial modes with 
$M=0.50\Msolar$,  $X=0.75, Z=0.02$ and enhanced
 iron ($f=10$); $\log L/M = 1.8$ was chosen to be
 representative of the EC14026 variables (see Fig.~\ref{sdbv_obs}).
It is evident that only the coolest stars are likely to pulsate in the
radial fundamental mode while many of the hottest can only excite the 
second or third harmonic. Some EC14026 stars lie beyond the blue-edge
for all of these models; a higher value of $f$ might have been more
appropriate, but the overall result is unaffected.  

Two exceptions are PG\,1605+072 and KPD\,1930+2752. The first has
evolved away from the extreme horizontal-branch and has a higher $L/M$ 
ratio (Fig.~\ref{sdbv_obs}). The second is tidally distorted by a very close
binary companion \citep{Bil00} and our classical radial models may be
inadequate. 

\begin{figure*}
\begin{center}
\epsfig{file=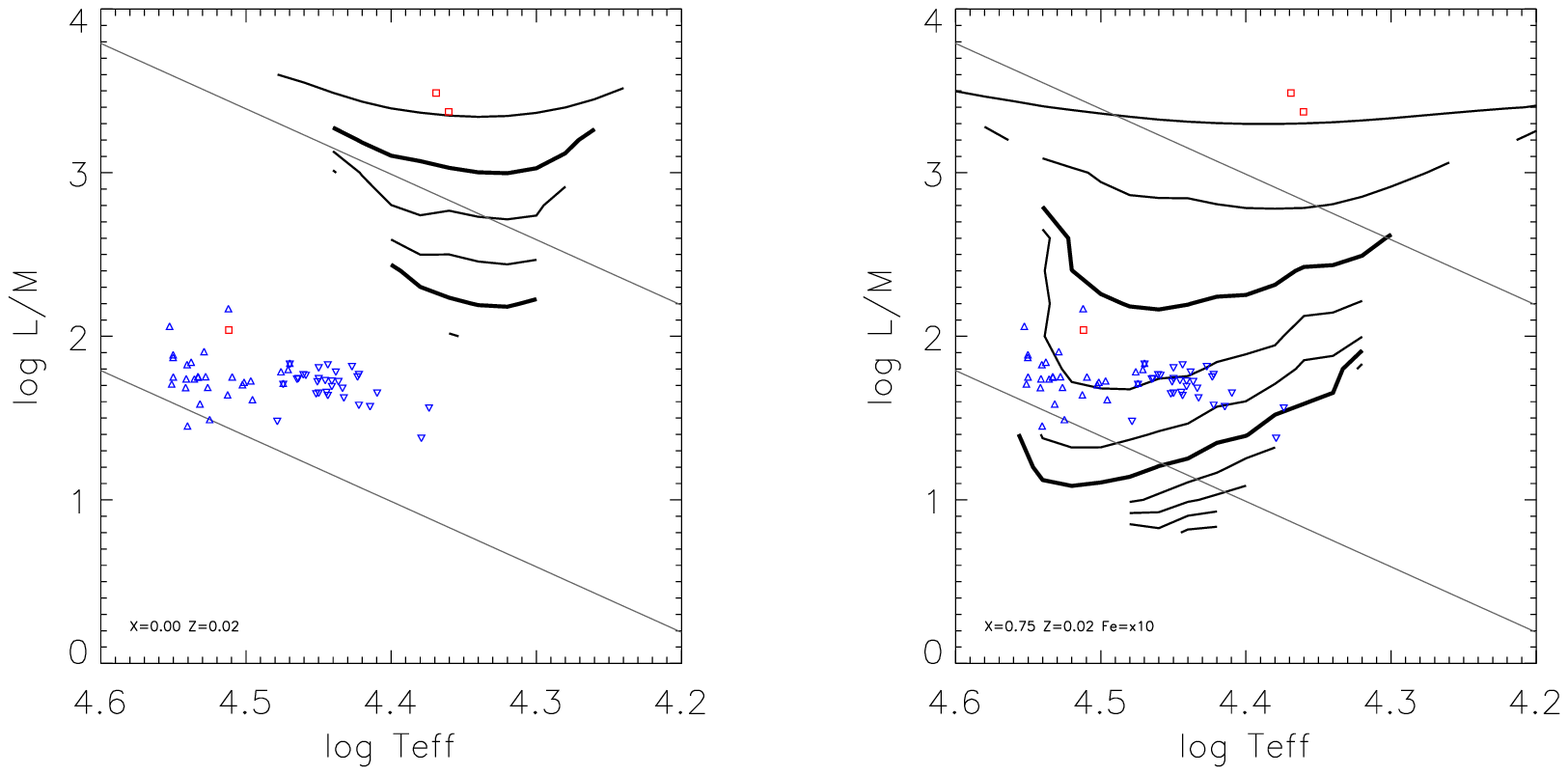,width=18cm,angle=0}
\caption[Growth Times]
{Contour plots representing minimum growth times for unstable modes
  in stars with $M=0.50\Msolar$ and for
  compositions representative of extreme helium stars (left)
  and EC14026 variables (right). Contours
  are separated by 1 dex, the bold lines representing
  $\log (-\omega_r/\omega_i)=3$ and 6, with growth times decreasing toward 
  higher $L/M$. 
  Other features as in Fig.~\ref{sdbv_obs}. 
}
\label{Growth}
\end{center}
\end{figure*}

\subsection*{Growth Rates}

An indication of the overall instability is given in Fig.~\ref{Growth}
where the exponential growth times are shown as contour plots for compositions
representative of extreme helium and EC14026 variables. The growth
times are represented by the dimensionless quantity
$-\omega_r/\omega_i$. 
Divide by $2\pi$ to find the growth
time in pulsation cycles, and then multiply by 
pulsation period $P$ to obtain the physical growth time $\tau_g$. Thus, for the EC14026 stars with
$M=0.50\Msolar$ and $\log L/M\sim1.7$, Fig.~\ref{Periods} indicates $\log
P/{\rm s}\sim 2$ and Fig.~\ref{Growth} indicates  $\log
(-\omega_r/\omega_i)\sim4$,
yielding $\log \tau_g/{\rm s}\sim5$. For the extreme
helium stars, the same exercise also gives $\log \tau_g/{\rm
  s}\sim5$. With growth times $\sim$ one day, these modes are highly unstable.

\begin{figure*}
\begin{center}
\epsfig{file=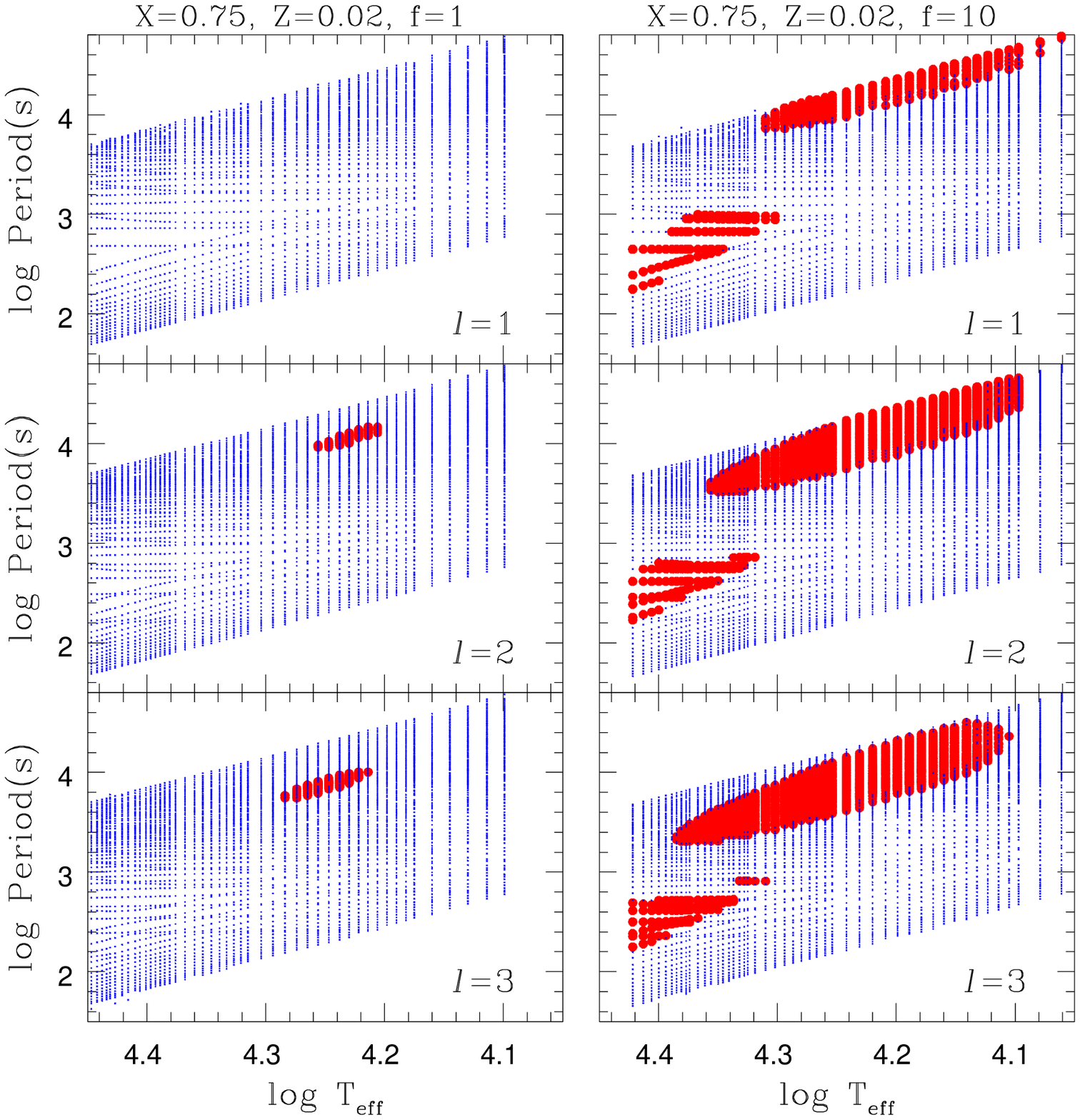,width=18cm,angle=0}
\caption[Non-radial stability analysis: $f=1$]
{Periods of modes due to Fe-bump instability for
  ZAHB stars with $M_{\rm c}=0.486\Msolar$, $X=0.75$ and $Z_0=0.02$, and
  $M_{rm e}=$ increasing from right to left. The left hand column
  shows models with no additional iron enhancement ($f=1$), the right
  hand column shows models with iron increased by a factor $f=10$. Stable modes are marked as
  (blue) dots, unstable modes are marked as filled (red) circles. 
  The chemical mixture with $f=1$ corresponds to that in the
  first column, second row of Figs.~\ref{Fe-stable} and
  \ref{Fe-modes},  while that with $f=10$ corresponds to that in the
  third column, second row of Figs.~\ref{Fe-stable} and \ref{Fe-modes}.
}
\label{nrad_f0110}
\end{center}
\end{figure*}

\begin{figure}
\begin{center}
\epsfig{file=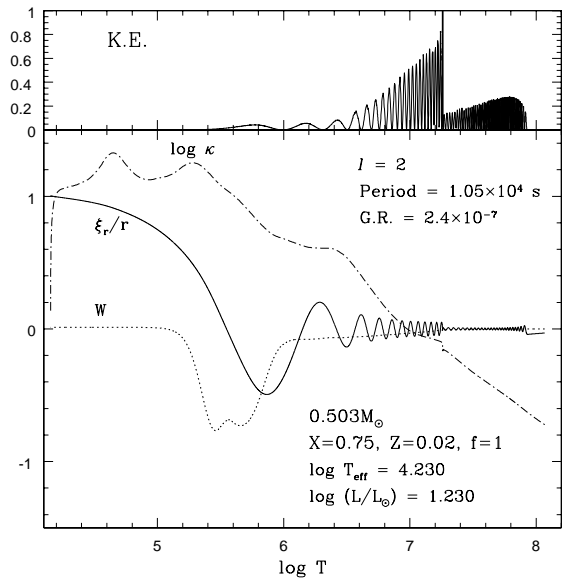,width=0.49\textwidth,angle=0}
\caption[]
{The bottom panel shows the radial displacement $\xi_r$ and  work $W$ of 
an unstable high-order g-mode of $l=2$, and opacity $\kappa$ as a function 
of the temperature. 
In a driving (damping) zone, $dW/dr > 0$ ($< 0$), and the mode is unstable 
if $W>0$ at the surface.
The top panel shows the distribution of the kinetic energy of the mode. 
A sudden change in the kinetic energy and in the amplitude of the spatial oscillation
at $\log T \approx 7.25$, where the hydrogen abundance changes steeply.
The convective core, where the g-modes are (spatially) evanescent, extends
to $\log T = 7.934$. G.R. is the growth rate $-\omega_i/\omega_r$. 
}
\label{work}
\end{center}
\end{figure}

\begin{figure*}
\begin{center}
\epsfig{file=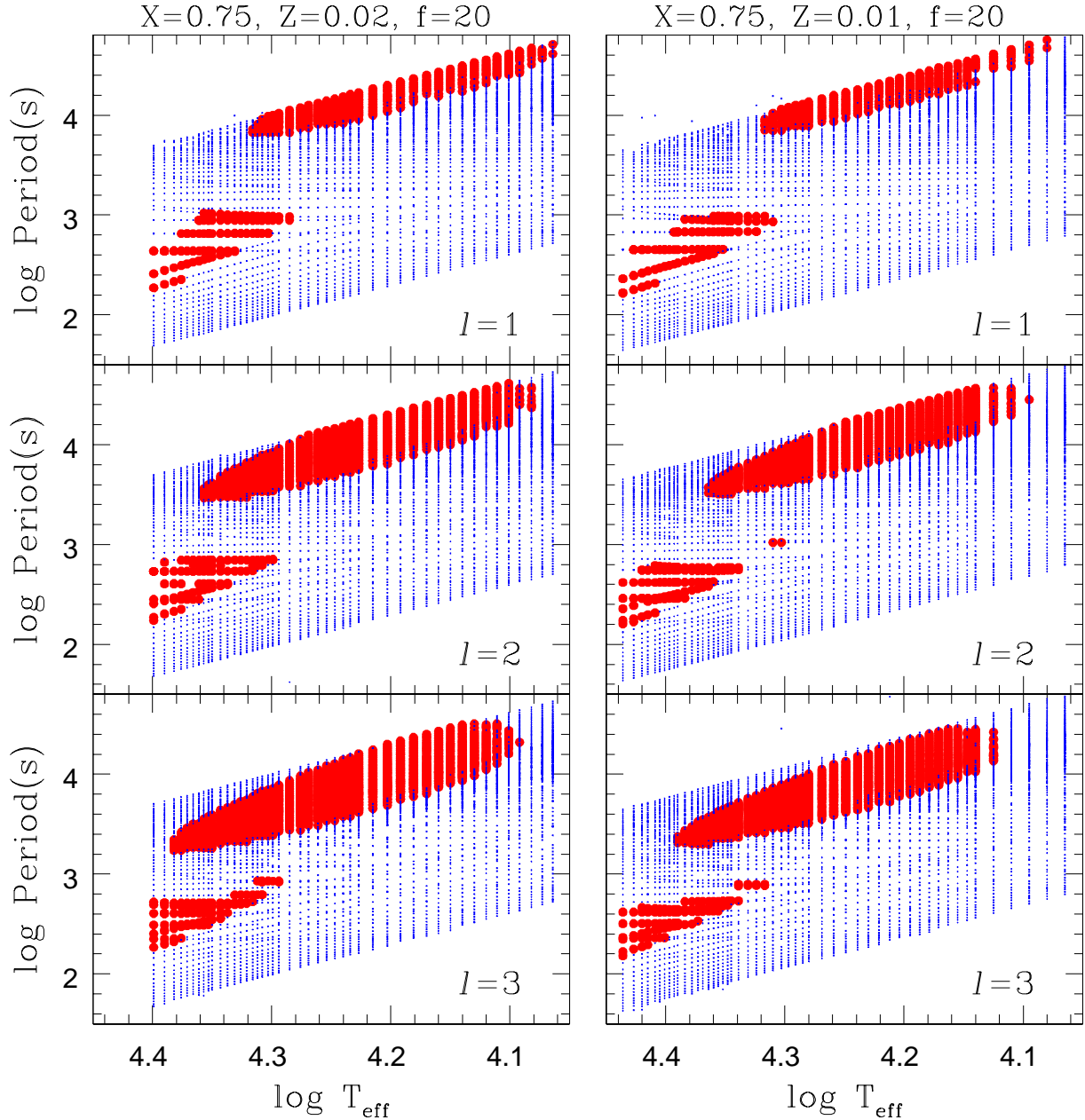,width=18cm,angle=0}
\caption[Non-radial stability analysis: $f=10$]
{As Fig.~\ref{nrad_f0110}, for models with 
  $M_{\rm c}=0.486\Msolar$, $X=0.75$ and iron enhancement $f=20$, 
  but for $Z=0.02$ (left) and 0.01(right), demonstrating the 
  shift of the g-mode blue edge with metallicity $Z$. 
}
\label{nrad_z0102}
\end{center}
\end{figure*}

\begin{figure*}
\begin{center}
\epsfig{file=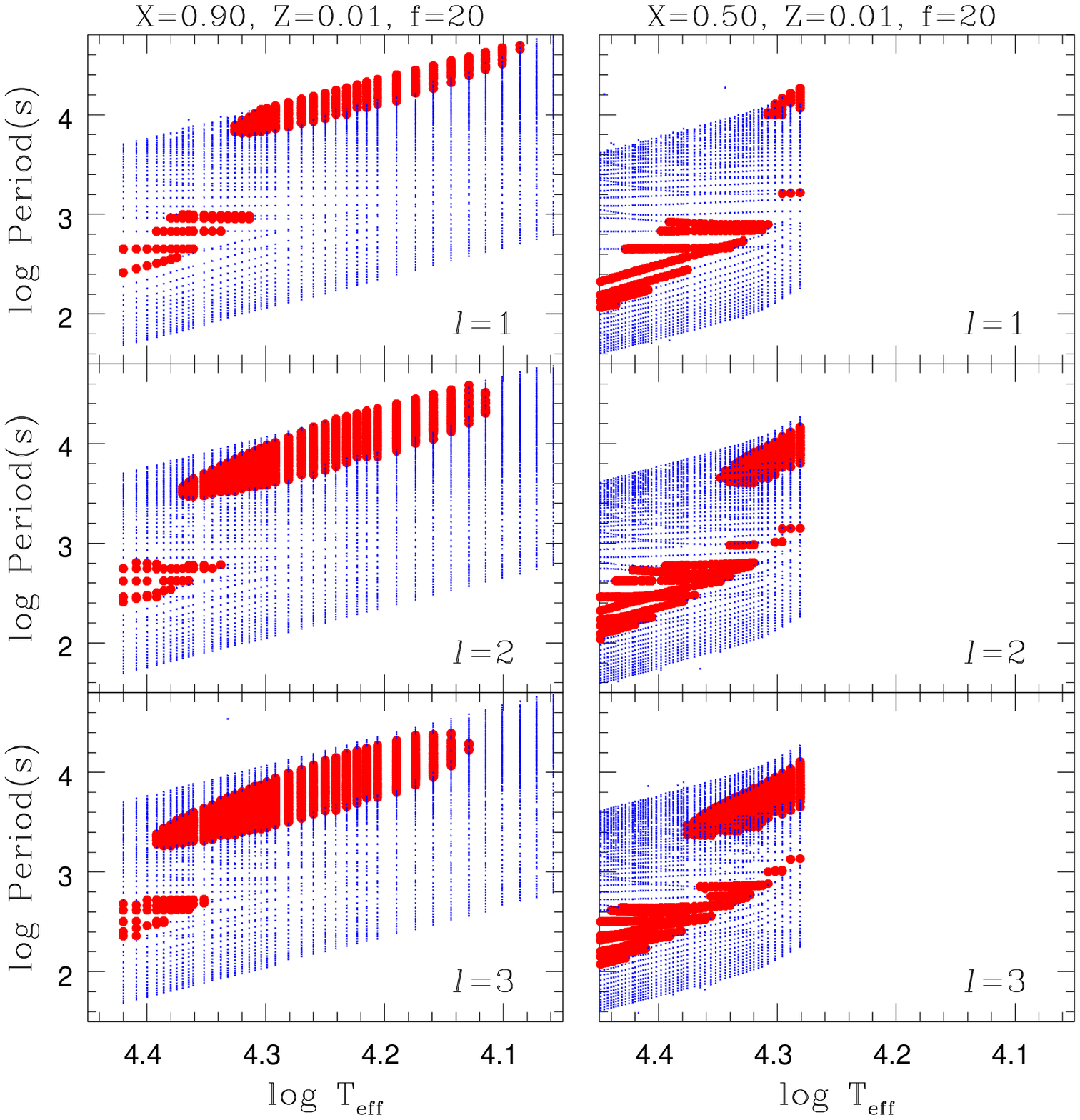,width=18cm,angle=0}
\caption[Non-radial stability analysis: $f=10$]
{As Fig.~\ref{nrad_f0110}, for models with 
  $M_{\rm c}=0.486\Msolar$, $Z=0.02$ and iron enhancement $f=20$, 
  but for $X=0.90$ (left) and 0.50(right), demonstrating the 
  shift of the g-mode blue edge with hydrogen fraction $X$. 
}
\label{nrad_x9050}
\end{center}
\end{figure*}

\section{Non-radial modes}        
\label{g-modes}

\subsection*{Zero-age extreme horizontal-branch stars}

The non-radial modes most usually observed in stars may be divided
into pressure or p-modes and gravity or g-modes. Pressure modes are
trapped in the outer stellar envelope and may be
thought of as a three dimensional generalization of the radial modes 
discussed in the previous section. Therefore their stability has
essentially been discussed already, although the precise periods of 
p-modes with spherical degree $l>0$ will differ from the
corresponding radial ($l=0$) values. 

Gravity modes propagate most strongly in the deep interior of the
star.
Consequently complete stellar models including both core and envelope are required. 
Following previous work \citep{Fon03} we have assumed that the PG1716
variables are 
extreme horizontal branch stars and have 
thus constructed series of ``zero-age'' horizontal-branch models
having a helium-burning core with a mass of $M_{\rm c}$
and with a hydrogen-rich envelope with mass $M_{\rm e}$ ranging from
 0.0003 to 0.0034 \Msolar. 
We have considered two cases of $M_{\rm c}$; $0.476$ and $0.486\Msolar$,
The surface layers are characterized by mass
fractions of hydrogen $X$ and base metallicity $Z_0$, with iron again augmented by
a factor $f$. We have considered sequences with $X=0.90, 0.75$ and
0.5, with $Z_0=0.01$ and 0.02, and with $f=1, 10$ and 20. 
The hydrogen abundance $X_{\rm tr}$ at the interface between a
hydrogen-rich envelope 
and a helium core is assumed to be given as
\begin{equation}
X_{\rm tr}(M_r) = 0.5X(1+\sin\theta)
\end{equation} 
with
\begin{equation}
\theta\equiv \pi [(M_r - M_c)/\Delta M -1],
\end{equation}
where $M_r$ is the mass coordinate within the transition layer,
and $\Delta M$, which is assumed to be $2\times10^{-4}\Msolar$, is the mass of the transition layer.

The high-temperature end of this sequence corresponds with the locus of
the short-period EC14026 variables. PG1716 variables are to be found
at lower temperatures, with $25\,000 \lesssim T_{\rm eff}/{\rm K} \lesssim 29\,000
$. 

We have carried out a stability analysis for each of these models
testing the stability 
of {\it non-radial} p- and g-modes spherical degree $l=1,\ldots,4$. 
The frequency range considered is 
$0.2 \le \omega \le 20$, where $\omega$ is the angular frequency of pulsation
normalized by $\sqrt{GM/R^3}$ with $G$ being the gravitational constant. 
The stability of non-radial modes is found to be insensitive to $M_{\rm c}$.
We show results for $M_{\rm c}=0.486\Msolar$. 
To allow for the likelihood that most horizontal branch
  stars have evolved away from their zero-age structure, we will
  demonstrate later that their principal stability characteristics survive
  through to the end of core helium burning. 

As anticipated, we find no unstable p-modes with $l>0$, for any models
with normal iron abundance ($f=1$). Like the radial modes, these are 
only excited with significant iron enhancements. Unexpectedly, 
we {\it do} find excited g-modes $l>1$ in ZAHB stars with $Z=0.02$ and
no iron enhancement over a small interval 
$ 16\,000 < T_{\rm eff}/{\rm K} < 20\,000 $, with $l=2$ modes being
excited in higher radial-orders and at lower $ T_{\rm eff}$ than $l=3$
and $l=4$ modes (Fig.~\ref{nrad_f0110}).

Some properties of an excited g-mode of $l=2$ are shown in Fig.~\ref{work},
where the solid line in the bottom panel is the radial displacement of pulsation divided
by the distance from the center $\xi_r/r$, 
the dotted line is the work $W$, and the top panel shows the distribution of the
kinetic energy of the mode. If a pulsation mode is excited (unstable), $W>0$ at the surface.
There is an excitation zone ($dW/dr > 0$) in a temperature range of
$5.1\la \log T \la 5.4$ due to the Fe-opacity bump and a damping zone ($dW/dr < 0$)
just below. 
This g-mode is marginally unstable because the driving effect is not very strong with no
enhancement of iron abundance.
In such a case, a reflection of the wave at the interface 
between the core and hydrogen-rich envelope plays an important r\^ole.
Fig.~\ref{work} shows that
the kinetic energy and the amplitude of the mode suddenly decrease towards the center
at $\log T \sim 7.25$, where the hydrogen abundance changes steeply, indicating a partial
reflection of the mode to occur there.
A sudden change in the mean molecular weight at the interface requires a sudden change in 
the gas density, which reflects g-modes partially.
The reflection reduces the amplitude in the core and hence reduces radiative damping
there.
The strength of the reflection depends on the phase of the spatial oscillation 
at the interface, and hence only some of the g-modes are excited in the optimal
period range for the Fe-bump mechanism. 

One model for which  g-modes are excited on the ZAHB was evolved
through its complete horizontal-branch evolution. The g-modes
continue to be excited throughout HB evolution, primarily
because, with no hydrogen-burning shell, the basic structure of the envelope
and the characteristics of the core-shell reflection surface are not altered.

The above result indicates that g-mode pulsations can be driven in certain blue
horizontal branch stars irrespective of whether chemical
stratification enhances the iron-abundance in the driving zone. 
This conjecture will require observational verification, which 
will constrain theoretical models for the blue horizontal branch stars. 
Moreover, it implies that g-modes in these cooler stars are more
likely to be excited with modest iron enhancements than the p-modes. 
The growth times for the ZAHB models are relatively long ($\sim 100$
y, Fig.~\ref{work}), but are reduced by factors of $\sim10$  during HB
evolution. 

However, in general, unstable modes (both p- and g-) are only recovered
with a significant enhancement of iron. Indeed, with 
$f\sim10$, we always find unstable modes of either p- or g-type 
down to $T_{\rm eff}\sim13\,000 {\rm K}$ (Fig.~\ref{nrad_f0110}). 
In these models, the reflection at the core boundary is not important 
in the excitation of g-modes, and most modes are excited in
an optimal range of periods. The growth times for these modes are 
$\sim 1 - 10$ y.
The locus of the p-mode
instability follows that of the radial modes as a function
of $X$, $Z$ and $f$.

In our models, with $f\geq10$, the high temperature limit of the unstable
g-modes always overlaps the low-temperature limit of the unstable 
p-modes of the same spherical degree. While this overlap is marginal for
$l=1$ modes, it increases substantially with $l$. This result has an important
corollary which states that if a star has an effective temperature
such that it is in the p- and g-mode instability overlap (p/g overlap)
and
if p-mode oscillations are observed, {\it i.e.} it is an EC14026
variable, then the iron enhancement necessary to drive the p-modes will
also be sufficient to drive the g-modes, {\it i.e.} it is also likely
to be a PG1716 variable. 

This is true regardless of the fact that we have used a global iron
enhancement rather than a stratified model. The reason is that the 
driving for both p- and g-modes comes from the same part of the star,
so that for a star of the right effective temperature, it may actually 
be impossible to have p-mode instability without g-mode instability. 
Indeed, the stars Balloon 090100001 and HS 0702+6043 lie on the
boundary between the observed EC14026 and PG1716 variables and exhibit
both types of pulsational behaviour \citep{Sch06,Ore05}.

However, there is a problem. The high temperature limit of the unstable
g-modes occurs around $T_{\rm eff} \sim 20\,000$ K for 
$l=1$ modes, increasing to $\sim 25\,000$ K for $l=4$. The
observed high temperature limit for PG1716 variables is  $\sim
29\,000$ K and it is unlikely that $l>2$. 

We have investigated g-model stability for ZAHB 
models having a range of chemical composition. Figs.~\ref{nrad_z0102}
and \ref{nrad_x9050} demonstrate the effects of varying the overall
metallicity $Z$ or the overall hydrogen concentration $X$ whilst
maintaining a high iron concentration ($f=10$). Increasing $X$ 
(equivalent to reducing the helium fraction) has the effect of a modest
increase in the effective temperature of the g-mode blue
edge. Similarly, reducing $Z$ while conserving the iron  abundance
(Fig.~\ref{nrad_z0102}) shifts the blue edge of g-modes blue-ward a little.
These effects can be understood from equation (\ref{eq_tauth})
However, in neither case is a substantial change
in the envelope composition capable of shifting the g-mode 
blue-edge to its observed location. 

Figs.~\ref{nrad_f0110} to 
\ref{nrad_x9050} exclude results for $l=4$. The $l=4$ blue edge is 
always bluer than the $l=3$ blue edge by $\sim 0.01$ dex.
In addition to the lower likelihood of observing $l=4$ modes, 
this is again insufficient to account for the observed blue edge.

These results closely mirror those obtained previously \citep{Fon03}, where 
radiative levitation provides the necessary enhancement in the iron
abundance in the driving region. These authors found g-mode
instability with $l=3,4$ at effective temperatures up to 24\,000 K
($\log T_{\rm eff} \leq 4.38$), or with $l\leq8$ up to 
$\log T_{\rm eff} \leq 4.43$, still too cool to account for the
observed blue edge. The main difference is that our models
show unstable g-modes at $l=1$ and 2 where those of Fontaine et al. 
did not. One reason may be that our model envelopes have a 
homogeneous chemical composition with an artificial iron enhancement,
whereas the earlier models have a stratified composition with iron
enhanced in critical layers by a well-understood mechanism. 

We have found that the effective temperature at the blue edge of g-modes
hardly changes when the degree of iron enrichment increases as long as
$f\ge 10$.
This indicates that to have a blue edge of g-modes consistent with the observation,
we have to change something in the structure of models.

\begin{figure}
\begin{center}
\epsfig{file=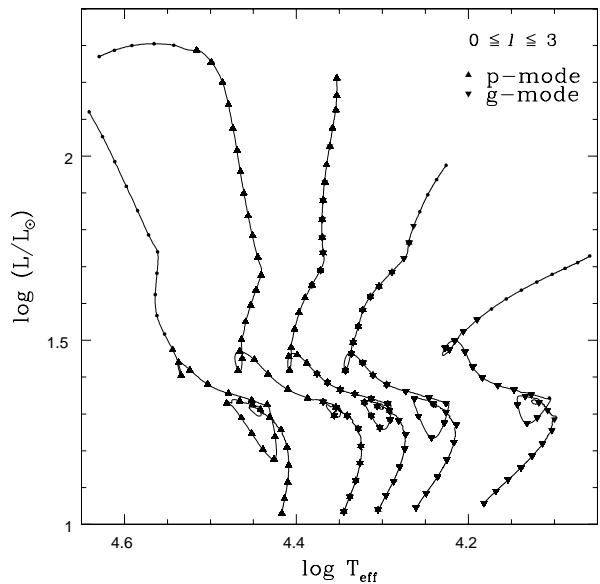,width=9cm,angle=0}
\caption[Non-radial stability and evolution]
{Evolution tracks for horizontal branch stars with $M_{\rm
    c}=0.476\Msolar$, $X=0.75$, $Z_0=0.02$ and enhanced iron abundance
  $f=10$. Total masses are $M=0.4763, 0.478, 0.480, 0.483$ and $0.490
  \Msolar$. Models having either unstable p-  or
  g-modes  with $l<4$ are  indicated by ($\blacktriangle$) and ($\blacktriangledown$), 
  respectively. Small dots indicate models in which neither is excited.
   Note that models having both unstable p- and g-modes appear as
  ``stars''. 
}
\label{nrad_evol}
\end{center}
\end{figure}

\subsection*{Evolved horizontal-branch stars}
\label{evolve} 

One solution to the blue-edge question might be to consider the evolution of these stars. We
have computed the evolution of iron-enriched extended-horizontal 
branch models with  $X=0.75, Z_0=0.02, f=10$ and  $M_{\rm c}=0.476\Msolar$ through the 
completion of core helium-burning and beyond (Fig.~\ref{nrad_evol}). 
Note the ``breathing
pulses'' (small loops in the HR diagram) towards the end of core helium-burning. These 
are encountered in our models with artificially high iron
abundances. They are due to the increased core opacity and are not 
significant in models with a normal core composition. Semiconvection
is included following the description by \citet{Spr92}

We have carried
out a stability analysis for selected models along each evolutionary
sequence. 
Fig.~\ref{nrad_evol} shows those models where either
p-modes and/or g-modes are unstable for $l<4$. 
Comparing the instability temperature range of zero-age models for the same composition,
we see that stability of p-modes are hardly affected by the evolutionary change
of the stellar structure, while g-modes tend to be damped in evolved models
in which the central part is radiative.
There is a 
clear overlap region $4.30 < \log T_{\rm eff}/{\rm K} < 4.38$, 
where both p- and g-modes are unstable. This extends upwards in luminosity 
to include some post-horizontal-branch stars; it does not extend
blue-wards.  

The Brunt-V\"ais\"al\"a frequency becomes extremely large 
in the evolved non-degenerate radiative core.
This makes the wavelength of a g-mode very short, which in turn enhances the radiative 
damping and stabilizes the mode.
Still some g-modes are excited even in evolved models with a radiative core. 
For these models a partial reflection at the bottom of the helium-rich envelope reduces
the amplitude and hence radiative damping in the core.
Since only few g-modes are excited and the evolutionary speed is fast, 
these g-modes are not very important observationally.

\section{Conclusion}                
\label{conclusion}

Our original goals were to understand oscillations  
observed in the helium-rich subdwarf LS\,IV$-14^{\circ}116$, and in the PG1716
variables. 
We therefore carried out a broad review of Fe-bump driven pulsational instability in
low-mass stars\footnote{Oscillations in 
main-sequence stars were deliberately excluded from our study.}. 

By first considering radial modes, we have
demonstrated the essential r\^ole of chemical composition such that
instability increases with the contrast between the iron-bump opacity 
and other opacity sources. This increased contrast may be 
achieved either by increasing
the iron abundance, confirming earlier work by \citet{Cha01} or 
by reducing the hydrogen abundance \citep{Jef98}. At least one
of these is necessary to excite oscillations in all of the {\it
  low-mass} Fe-bump pulsators discovered to date. 

We have further demonstrated that the blue-edge for radial instability 
is affected by the mean molecular weight in the stellar envelope, so that
increasing the iron abundance alone provides a bluer instability region
than increasing the iron abundance in concert with all elements
heavier than helium. The former is required to explain the locus of the
EC14026 instability region, making the general assumption that the
properties of non-radial p-modes modes are closely linked to 
the corresponding radial mode of the same radial order. 

Furthermore, the blue-edge also depends on the radial order
of the oscillations, so that higher-order modes may be found in hotter
stars. By comparing theoretical and observed periods for EC14026 stars,
we have shown that low-order or fundamental mode radial oscillations 
are only likely to be seen in the coolest EC14026 stars and that the hottest stars
must oscillate in higher-order modes. 
 
However, we have been unable to explain the oscillations in the helium-rich 
star  LS\,IV$-14^{\circ}116$; the observed periods are simply too 
long compared with the $L/M$ ratio obtained from spectroscopy.
A lower surface gravity would definitely help. 

Considering non-radial pulsations, we have 
focused on g-mode instability and in particular on instability in
modes with spherical degree $l<4$. We have discovered a small 
g-mode instability island on the blue horizontal branch which 
does not require iron enhancement or hydrogen depletion. However, 
since it requires $Z=0.02$ and a partial reflection of the wave at the
interface between the hydrogen-rich envelope and the helium core, 
it will take some observational effort to
verify whether there are any real pulsating horizontal-branch stars
which correspond with these models. Effective temperatures are 
expected to be between 16\,000 and 20\,000 K, and pulsation periods
between 1 and 4 hours. 

With a factor of 10 enhancement of iron on a background metallicity
 $Z=0.02$,  a large instability region develops, even for $l=1$, which 
extends from $<13\,000$ to $\sim 24\,000$ K ($l=3$). With such iron
 enhancement there is always an overlap between p- and g-mode
 instability regions so that stars near the boundary would be expected
 to exhibit both modes simultaneously. 
Depleting the envelope hydrogen 
abundance tends to shift both the g-mode blue edge and the
radial/p-mode red edge to lower temperatures. 

Many of these results have been demonstrated individually 
\citep{Sai93,Jef98,Cha96,Fon03}; this investigation has 
placed them in a more general context, as well as delivering some new
results. Of these, the r\^ole that mean molecular weight 
plays in determining the instability boundaries for radial and p-mode 
oscillations has implications for understanding the oscillations in
EC14026 variables. We have made no effort to justify the adopted
abundances on physical grounds, we have simply sought parametric
solutions which satisfy the observations. Radiative levitation 
\citep{Cha95} is known to operate in extreme horizontal-branch stars
and provides a natural explanation for the required iron
enhancements. The fact that it operates selectively accords 
well with our deduction that only iron should be enhanced. 

Our discovery that g-mode oscillations may be excited in 
blue horizontal branch stars with envelopes of ``normal'' composition
has two-fold consequences. In addition to the observational
question already posed, it may be less hard to 
excite g-mode pulsations in PG1716 stars than hitherto supposed; maybe
we don't have to work so hard to find the necessary chemical structure
in the stellar envelope as we do in the case of EC14026 variables.

Nonetheless, we have been unable to resolve the problem of why the observed
boundary between the EC14026 and PG1716 stars occurs at $\sim 29\,000$
K. The theoretical p-mode/g-mode boundary remains persistently at 
 $\sim 24\,000$ K over a wide range of envelope compositions. It is
possible that a sharp discontinuity in composition immediately below
the iron bump could help; this and other experiments with stratified 
envelopes remain to be investigated.

\section*{Acknowledgment}

Travel support for this collaborative project was provided through 
PPARC grant PPA/G/S/2002/00546.

\end{document}